\documentclass[12pt]{article}
\usepackage{amsmath}
\usepackage{amssymb}
\usepackage{amsthm}
\usepackage{graphicx}
\usepackage{cite}

\title{The Saga of KPR: Theoretical and Experimental developments}
\author{Kiran Sharma\footnote{Email address: kiran34\_sit@jnu.ac.in,
School of Computational and Integrative Sciences, Jawaharlal Nehru University, New Delhi-110067, India}, 
Anamika\footnote{Email address: anamika@jnu.ac.in,
School of Computational and Integrative Sciences, Jawaharlal Nehru University, New Delhi-110067, India}, 
\\ 
Anindya S. Chakrabarti\footnote{Email address: anindyac@iima.ac.in,
Economics area, Indian Institute of Management, Ahmedabad, Gujarat, India}, 
Anirban Chakraborti\footnote{Email address: anirban@jnu.ac.in,
School of Computational and Integrative Sciences, Jawaharlal Nehru University, New Delhi-110067, India},
\\Sujoy Chakravarty\footnote{Email address: sujoyc@gmail.com,
School of Social Sciences, Jawaharlal Nehru University, New Delhi-110067, India}}
\date{\today}

\begin{document}

\maketitle

\begin{abstract}
\noindent In this article, we present a brief narration of the origin and the overview of the recent developments done on the {\it Kolkata Paise Restaurant} (KPR) problem, which can serve as a prototype for a broader class of resource allocation problems in the presence of a large number of competing agents, typically studied using coordination and anti-coordination games. We discuss the KPR and its several extensions, as well as its applications in many economic and social phenomena. We end the article with some discussions on our ongoing experimental analysis of the same problem. We demonstrate that this provides an interesting picture of how people analyze {\it complex} situations, and design their strategies or react to them. 
\end{abstract}

\section{Introduction} 
\noindent  Chakrabarti et al. \cite{chakrabarti2009, Ghosh2010} introduced the {\it Kolkata Paise Restaurant} problem (KPR henceforth), which was a multi-agent, multi-choice game. The original motivation of the problem was
to introduce the statistical analysis of a large-scale coordination game. The game theoretic solution concepts were introduced and it was shown that the agents (bots) could attain surprising level of coordination using some simple heuristics, even without any global coordinator. This approach was parallel to the one followed by Brian Arthur in his famous paper on El Farol bar problem (EFB henceforth) \cite{Arthur1994a}, which later gave birth to the {\it minority game} (MG henceforth) formulated by the physicists \cite{mgpapers1}. 

The essential idea of Arthur was to show that heuristics based on induction, can be very useful for solving a game theoretic scenario. The KPR problem can be viewed as simply a straight-forward generalization of the EFB problem, yet mathematically tractable. However, as we would mention in this paper, notwithstanding the huge amount of research done on the MG problem \cite{Challet2004}, KPR provides a bigger laboratory to study many ``complex'' emergent phenomena. In recent times, it has been also used for operations research problems, which have enormous practical applications. 
For example, the taxi services (like Uber and Ola) solve a bilateral matching problem and KPR provides a starting point for analyzing such games \cite{Iyer17}.
For the sake of completeness, we  state only two versions of the KPR problem here, though there are good resources for a much more complete description and discussion. Interested readers
can refer to Ref. \cite{Chakraborti2015,Chakraborti2010} and \cite{KPRbook17}. 

We can define two versions of the KPR, one without ranking and the second with ranking. The first one is simpler and closer to a standard (anti-) coordination game, and so we begin with the one without ranking.
\begin{description}
\item[Game 1] Consider a set of $n$ players and a set of $m$ identical restaurants ($n=m$ is the most interesting case). Each player has access to its own set of information, which can be null (in principle), and has to secure access to one restaurant (win). Each of the players has to decide which restaurant to go, and if more than one player arrives at 
any restaurant, then only one of them is chosen \textit{randomly} with equal probability. 
\item [Game 2] Identical to the above, except that the restaurants are now \textit{ranked}, which is commonly agreed upon and known to all the players. 
\end{description}

For the first game (and with the symmetric case, $n=m$), the Nash equilibrium could be simply achieved by assigning each of the agents to one restaurant only, such that all $m$ restaurants are allocated and there are $n$ winners. In the second case, it is certainly more complicated -- imagine all the $n$ agents being assigned to the $m$ restaurants one-to-one, and solving the KPR problem. Since the restaurants are now ranked, there would be incentives for the person being assigned to lower rank restaurants to then deviate and choose upper ranked restaurants. This type of cases have been considered in details \cite{KPRbook17}, and Chakrabarti et al.\cite{chakrabarti2009} provides conditions under which there exist pure strategy Nash equilibria. Below, we briefly discuss the different facets of the KPR.

\section{KPR at the Crossroads of Disciplines}

There exists significant amount of research done on coordination problems in multiple disciplines. Computer scientists have worked on parallel processing of
jobs and tried to solve allocation problems by generating efficient algorithms. Statistical physics literature has dealt with traffic models, which again relate to
individual choices leading to congestion and hence inefficiency. Obviously, from a game theoretic perspective, the apparently different scenario can be understood essentially as one {\it congestion} game. There are many other possible applications and interpretations of the game. Hence, we consider the KPR problem to represent a paradigm rather than a single, specific problem. In the following, we describe some directions the KPR problem has been studied and the links with existing literature across disciplines.

\subsection{Heuristics to Solve a Complex Coordination Problem}
A major part of the research on KPR has been to model it as a complex coordination problem and show two things: Firstly, simple and local heuristics can be useful for solving global optimization problem. Secondly, it can provide insights into how effective the different heuristics are to solve the problem. Imagine that this is a resource competition game and each successful match leads to reproductive success. Then the strategy pool would start evolving favoring more successful strategies. Chakrabarti and Ghosh (see Refs. \cite{Chak_Gho_17a} and \cite{Chak_Gho_17b}) considered a sequence of
heuristics and also introduced limited form of learning and analyzed the relative successes. It turned out that simple rules of reinforcement learning with little use of any global or local information of the agents, could actually solve the allocation problem with very high efficiency.

\subsection{Attaining Distributed Coordination}

An important problem in designing efficient solvers is to find a way to distribute the load across different processors to attain parallel processing.
One way to think about it, is to consider a central coordinator that would assign tasks across processors. However, in a multi-agent system
with a very large number of agents, it may be prohibitively expensive to program such task (or resource) allocation \textit{a priori} for all possible contingencies.
Therefore, we may like to have {\it intelligent} agents who can strategize and discover the way to divide tasks by themselves. 

Cigler and Faltings \cite{Cigler_13} considered this approach and suggested that a correlated equilibrium (proposed in the game theory literature by Aumann)
could solve this problem; but they also recognized that such a device (which induces anti-coordination) basically acts as the central coordinator, and so this idea did not prove to be very useful. So they further proposed some learning rules, which are useful for solving this kind of problems.
Hansen and Giauque \cite{Hansen1986} proposed dynamic programming algorithm as an alternate for task allocation problems. There are many other papers, which addressed similar questions
(see for example, Refs. \cite{Chu1980}, \cite{Doty1982}, \cite{Galstyan_05}).

\subsection{Connections to Game Theory}

The KPR is essentially a game theoretic problem, and so it is least surprising that economists would be interested in a similar paradigm. The original formulation of the KPR problem assumed that not only the agents have a ranking over the restaurants, but also the ranking is identically agreed upon and known to all agents. More explicitly, each and every agent
agrees on which one is the better restaurant in any given pair of restaurants. A more complex version of the same problem was already known as the
{\it Stable Marriage Problem}, where one assumed that there were two groups of members (identical in number), say men and women, who would marry each other\footnote{Same-sex marriage is not considered in the present context.}. Each man had a ranking of the women, and vice versa. The problem was to find the matches across members of the two groups such that there would be no two members who would rather break up and marry each other.

Gale and Shapley in 1962 provided an algorithm to solve this problem \cite{GaleShapley1962}. This problem has related variants like assignment problem, where
a number of agents have to be matched with the same number of tasks such that the matching is one-to-one. The catch is that every match has a cost of implementation and so the problem is to minimize the total cost of all matches.
Such problems are of immense importance from a practical point of view. Roth and Pernson \cite{Roth99} provided a detailed algorithm along with applications
for solving such allocation problems.

It is clear that although KPR deals with an allocation problem, there exist significant differences with respect to the above-mentioned approaches. One of the crucial differences is that
in the above problems, it is assumed that there exists a central coordinator who allocates the task, whereas in KPR there exists none. So, we now move from the discussion on a {\it central planner} to a {\it decentralized} outcome.

From the game theoretic perspective, an important class of games known as {\it congestion games}, are close to the KPR problem. This class of games also assumes that
players can choose any one among multiple options (roads) and if the same option is chosen by more than one player (leading to congestion) then that is bad for all of them. This can be mapped to the KPR problem (without ranking) upon simplification.

The closest problem was stated by Grenager et al. \cite{Grenager_02}, as well as Alpern and Reyniers \cite{Alpern_02} who called it {\it dispersion game}. However, KPR with ranking has a more general structure. Below, we discuss the connection of the KPR with agent-based modelling literature and in particular, with the minority game literature.

\subsection{Agent-based Models and Connections to MG}

A variant of the EFB problem was proposed and analyzed by Challet and Zhang, which later on gained considerable popularity as the minority game (MG) \cite{mgbook,mgpapers1,mgpapers2,mg1,mg2,rev-chakraborti2011}.  
Structure-wise, MG is a $N$-agents, 2-choice repeated game.
Choices are defined over two restaurants are considered and the agents on the less-crowded side are considered to be the 
winners while the rest of them are declared losers. 
The primary aim of introducing the MG  was to serve as simple a model of a financial market. Two choices are identified as `buy' or `sell' options executed by the trades. More buyers than sellers in any day would imply gain (positive payoff) for sellers and the opposite when there were more sellers than buyers. The total payoff of all agents could be treated as equivalent to the excess demand of the market, and be considered as the price return (directly proportional to  the excess demand) of the market \cite{mgbook}. 

Traders would learn from their own history of the game and alter their optimization strategy as the need be. The fluctuations of the players using different strategies in the steady state 
reflects the inherently dynamical and adaptive nature of the game. 
Agent-based modeling and Monte Carlo simulations, along with tools borrowed from statistical physics were used to show that the variability of fluctuation can be minimized in terms of memory size and number of players.

Physicists and economists became interested in related problems like the crowd versus anti-crowd movement of agents, broadly in the context of opinion formation.  MG has served as prototype model for evolving and adaptive systems. Sysi-Aho et al. \cite{sysiaho1,sysiaho2,sysiaho3,sysiaho4}, in a sequence of papers, studied a modified version of
the MG where the players were able to choose newer strategies
using mechanisms of genetic crossover depending on their relative performances. 
This modification was useful to make the agent-based models applicable to many real-world systems especially with heterogeneous agents and non-linear interactions.

\section{Could laboratory experiments provide new insight?}

Numerous laboratory experiments have tested whether or not human agents play Nash equilibrium strategies in various 
different games, both one-shot and sequential. Some one-shot games with unique pure strategy Nash equilibria that have been tested in the 
laboratory include the prisoners' dilemma \cite{Flood_58,Dawes_80}, voluntary contributions mechanism (VCM) games 
\cite{Isaac_88}\footnote{See Refs. \cite{Sally_95} and \cite{Zelmer_03} for meta-analyses of prisoners' dilemma and linear public goods games, 
respectively.}, auctions \cite{Kagel_08}, Bertrand pricing game \cite{Dufwenberg_00} and the travelers' dilemma \cite{Capra_99}. 
A few sequential games, including the ultimatum bargaining game \cite{Guth_82,Forsythe_94} and the trust game \cite{Berg_95}, have been studied in the laboratory. The results from these studies broadly establish that human players are not as \textit{hyper-rational} as predicted by theory, and the laboratory outcome is often not the Nash equilibrium\footnote{See Ref. \cite{Johnson_11} for a meta-analysis of trust game experiments.}.  In almost all studies, \textit{bounded rationality} as well as concerns for \textit{equality}, affect human decision-making. Stahl and Wilson \cite{Stahl_95} introduced level-$k$ rationality model that attempts to model different levels of rationality that a human player in a one-shot game may display. Models of other-regarding preferences that have been informed by results from laboratory experiments include those by Rabin
\cite{Rabin_93}, Fehr and Schmidt \cite{Fehr_99}, and Bolton and Ockenfels\cite{Bolton_00}.

The class of games that is particularly relevant for our study is coordination games. These are modeled in their normal form as one shot games with multiple Nash equilibria that can be Pareto rankable \cite{Bowles_04} or non-Pareto rankable \cite{Schelling_60,Mehta_94}. In such games, an empirical question that interests game theorists and experimental economists is whether agents manage to play the coordination outcome that realizes for them the highest level of joint profit, i.e., the welfare maximizing or efficient outcome.  Given that a refinement is necessary to identify the more likely ones from this set of multiple equilibria, game theorists and experimentalists have developed possible selection criteria of equilibria and tested their empirical validity. Broadly speaking, the selection criteria that have been suggested either rely on structural properties of games or salience in the form of naturally occurring psychological labels. An important criterion relying on the structural properties of games involves the ideas of payoff and risk dominance was introduced by Harsanyi and Selten \cite{Harsanyi_88}. The theory of hierarchical decision-making that relies on the existence of focal points was first posited by Schelling \cite{Schelling_60} and tested by Mehta et al. \cite{Mehta_94}. Accordingly, naturally occurring labels that are of common knowledge can allow for greater coordination, particularly, in situations where there exist many non-Pareto rankable Nash equilibria.

Important experimental studies of coordination games with multiple Pareto ranked Nash equilibria includes the work of Van Huyck et al. \cite{Huyck_90}, who study the minimum effort coordination game with seven Pareto rankable Nash equilibria with 14 to 16 subjects. They find on static repetition that subjects initially attempt to coordinate on the payoff dominant Nash equilibria but within a few periods only manage to 
converge to the risk dominant equilibrium. Cooper et al. \cite{Cooper_92} use a much simpler two player three strategy design, and in a similar fashion find that contrary to what Harsanyi and Selten \cite{Harsanyi_88} argue. Pareto dominance does not prove to be a robust predictor for equilibrium selection and dominated strategies are often not decision-irrelevant. Straub \cite{Straub_05} reaches a conclusion similar to Cooper et al. \cite{Cooper_92} by using simple two by two coordination games.

Strong path dependence is seen in another work by Huyck et al. \cite{Huyck_97} in a median-action ``continental divide'' game, where there are two Pareto ranked equilibria. If in the initial rounds of a finitely repeated game session, participants are close to the high payoff equilibrium, then they tend to remain close to this equilibrium over all the rounds. A similar pattern is seen for sessions where choices remain close to the lower payoff equilibrium for the whole session. This history dependence in statically repeated one-shot games opens up the possibility for applying lagged behavioural models such as the reinforcement learning model \cite{Erev_98} and the Experience Weighted Attraction (EWA) model \cite{Camerer_99}.

An important goal of our KPR experiment \cite{Kiran2017} is to test the various strategies that are employed by incentivized human subjects in the complex situation where a solution cannot be deduced. There are a number of points that we would like to see: (i) Whether or not these strategies move the cohort of players closer to the efficient outcome; (ii) If there is any effect of experience and information feedback on coordination to the efficient outcome; (iii) The usage of learning models in order to model the decisions taken in the experiment, and (iv) The effect of external mechanisms that can potentially increase coordination efficiency and lead to better utilization. 

In the experimental data collected so far, we do not see a significant effect of information and experience in increasing the utilization ratio. In our experiment, we broadly observe a few behavioral types: (a) the ‘noise traders’ -- who are reactive and change their choices very frequently, (b) the completely stable agents -- who stick to one strategy over all the periods, and (c) the intermediate type -- who change their  strategies a few times over the experiment, though not as much as the agents in case (a). We also check whether changes in strategy are influenced by the outcome in the previous period and find that going from failure to success in the previous period significantly lowers the likelihood of a player changing his or her strategy in the current period. The results do not exactly match with similar simulation studies that were conducted before, and hence provide new insights. We do need to conduct more such experiments and understand how human beings \textit{actually} strategize and adapt in ``complex'' situations.

\section{Summary and Outlook}

The KPR serves as a prototype model 
describing a general problem of competition over finite resources in a large population with a fully decentralized
decision process, i.e., without any communication between the competitors. The repeated interactions and the associated learning mechanism 
lead to emergence of collective dynamics in the form of fluctuations in resource utilization.
Additionally, in this competitive framework, the agents 
choose ot be in the minority. Hence, they need to act as differently as possible through adaptation and modification of strategies, which would entail heterogeneous and non-equilibrium dynamics such as those found in many complex systems. 
As described above, the KPR encapsulates multiple facets of complexity theory and connects multiple disciplines. We hope that in the near future, this game will serve as a cornerstone for studying global optimization problem from localized active {\it smart} agents, in both theory and experiment.

\section*{Acknowledgements}

AC and KS acknowledge the support by grant number BT/BI/03/004/2003(C) of Govt. of India, Ministry of Science and Technology, Department of Biotechnology, Bioinformatics division, University of Potential Excellence-II grant (Project ID-47) of JNU, New Delhi, and the DST-PURSE grant given to JNU by the Department of Science and Technology, Government of India. KS acknowledges the University Grants Commission (Ministry of Human Research Development, Govt. of India) for her senior research fellowship. ASC acknowledges the support by the institute grant (R\&P), IIM Ahmedabad. The authors thank the IT staff of SCIS, JNU, for their help in conducting the experiments.

\end{document}